\documentclass[12pt,a4paper]{article}

\usepackage{amsfonts}                   \usepackage{amssymb} 
\usepackage{amsmath}
\usepackage{a4wide}                     \hfuzz=10pt 
                      \def\ad{\mbox{ad}\,}

\def\c#1{{\cal #1}}                             
                    
\def\Dirac{{\raise0.09em\hbox{/}}\kern-0.69em D}
                
\def\ep{i\epsilon}                      
                                          
\def\kbar{{\mathchar'26\mkern-9muk}}    
\def\kb{i\kbar}

\def\lesssim{\mathrel{\hbox{\rlap{hbox{\lower8pt\hbox{$\sim$}}}\hbox{$<$}}}}
                                   
                    \def\sq{\hbox{\rlap{$\sqcap$}$\sqcup$}}  
                                     
\def\p{\partial}                                        
\def\tfrac #1#2{\textstyle{\frac{#1}{#2}}}
\def\dfrac #1#2{\displaystyle{\frac{#1}{#2}}} 	
\def\t#1{\tilde{#1}}

\def\k{\kern-.1em\mathbin{,}\kern-.1em}
 
\def\hk{\kern.12em\raise-1em\hbox{$\hat{\raise1em\hbox{,}}$}\kern.12em}

\newcounter{eg}                         \newtheorem{eg}{Example}[section]        
\def\beg{\begin{eg}\rm}                 \def\eeg{\hfill\sq\end{eg}}

\newcommand{\initiate}{\setcounter{equation}{0}}

\begin{document} 
\title{Noncommutative geometry of phase space}
\author{Maja Buri\'c $^{1}$\thanks{majab@phy.bg.ac.rs} 
                   \ \ 
        John Madore$^{2}$\thanks{madore@th.u-psud.fr} 
                   \\[15pt]$\strut^{1}$
        Faculty of Physics 
                   \\
        University of Belgrade, P.O. Box 368 
                   \\
        SR-11001 Belgrade 
                   \\[5pt]$\strut^{2}$
        Laboratoire de Physique Th\'eorique 
                   \\
        Universit\'e de Paris-Sud, B\^atiment 211 
                   \\
        F-91405 Orsay 
                   \\
       }

\date{}
\maketitle
\parindent 0pt 
\vskip25pt 

\begin{abstract}
  A version of noncommutative geometry is proposed which is based 
  on phase-space rather than position space. The momenta encode the
  information contained in the algebra of forms by a map which is the
  noncommutative extension of the duality between the tangent bundle
  and the cotangent bundle.
\end{abstract}

\thispagestyle{empty} \vfill 
\verb=$Id: ncp.tex,v 1.76 2011/07/06 02:36:58 madore Exp $=
\newpage
\tableofcontents \thispagestyle{empty} 
\newpage
\setlength{\parskip}{15pt plus5pt minus0pt}

\initiate
\section{Introduction}                \label{Iam}

It has been long conjectured~\cite{Pau56} that quantum fluctuations of
the gravitational field might soften the singularities which
appear in almost all solutions to the gravitational field equations
without source. We here assume that this role is to be played by the
effect of noncommutativity. We assume that is that noncommutativity is an
effective if but partial description of quantum fluctuations; it would
play in this respect the same role that thermodynamics plays with
respect to statistical physics. Noncommutative geometry to a large extent 
follows a path
previously taken by quantum mechanics with one essential difference;
there is no experimental motivation for it. We consider the subject of
interest in so far that it can be considered as an extension of the
theory of gravity which enables us to better understand the role of
the gravitational field as a  regulator for classical as well
as quantum singularities. 

Over the passed few years a noncommutative generalization of the
Cartan frame formalism has been developed~\cite{Mad89c,Mad00c} and
applied~\cite{BurGraMadZou06a,BurGraMadZou06,AscSteMadManZou07,
DouGraMadZou07}
with varying degrees of success to problems in gravitational physics,
notably to the possibility of blowing up~\cite{MacMadManZou03} the Big
Bang. One distinguishing feature of this formalism is that the field
equations are not derived from an action principle but rather from
constraints imposed on the frame arising from Jacobi identities. These
identities, we conjecture, fix the Ricci tensor. In particular
then the Einstein tensor is determined and the value it takes can be
interpreted as an effective source due to the existence of the
noncommutative structure. In the quasi-classical approximation this
can be more precisely interpreted as the energy of the Poisson
structure.  One of the motivations of this modification of the
 field equations is the further conjecture that curvature as
is usually defined in differential geometry cannot be used without
modification in the noncommutative generalization. It is hoped that
once the correct expression is found then the Ricci contraction will
yield an Einstein tensor which includes the Poisson energy.

The notation is drawn from previous publications~\cite{Mad89c}. A
tilde is used to designate  commutative geometry or a
classical limit of a noncommutative one.  We use the convention 
of distinguishing the frame
components of a vector from the natural components by a choice in
index: the Greek or Latin indices from the middle of the alphabet are 
coordinate and those from the beginning are frame indices.
 For example if $J^{\mu\nu}$ are the natural components of an antisymmetric
tensor then the frame components will be written as $J^{\alpha\beta}$. 
If $J^{\mu\nu}$ is a commutator this is
quite consistent provided we remain at the quasi-classical
approximation. We shall however apply the convention also when
describing the nonperturbative solutions in Section~\ref{np} in which
case it can be ambiguous.

\subsection{Algebraic approach}

Consider a smooth manifold $M$ with a moving frame
$\t{\theta}^\alpha$.  Let $\c{A}$ be a noncommutative deformation of
the algebra $\c{C}(M)$ of smooth functions on $M$ defined by a
symplectic structure $J$ and let $\theta^\alpha$ be a noncommutative
deformation of the moving frame. As we shall see, the connection 
on $\c{A}$  can be defined to satisfy both a left and right Leibniz rule, 
a condition which is  intimately connected with the existence of a 
reality condition; the metric can be defined to be a
bilinear map, a condition connected with locality.  The classical
limit of the geometry is thus naturally equipped with a linear
connection and a metric as well as with a Poisson structure.  Under
the assumptions which we shall impose the Poisson structure is
non-degenerate.  We shall be more precise about these extensions
below. We would like to be able to show that in the weak-field,
quasi-classical approximation they imply that the metric defined by
the frame cannot be arbitrary and that the Ricci tensor is fixed by
Jacobi identities.

Let $\c{A}$ be a noncommutative $*$-algebra generated by four
hermitian elements $x^\mu$ which satisfy the commutation relations:
\begin{equation}
[x^\mu, x^\nu] = i\kbar J^{\mu\nu}(x^\sigma).        \label{xx}
\end{equation}
As a measure of noncommutativity, and to recall the many parallels
with quantum mechanics, we use the symbol $\kbar$, which 
designates the square of a real number whose value could lie somewhere
between the Planck length and the proton radius $m_P^{-1}$. The value
becomes important when we consider perturbations. 
The $J^{\mu\nu}$ are of course restricted by Jacobi identities; we shall see
below that there are two other natural requirements which also
restrict them.

Let $L$ be a macroscopic length scale. In the Schwarzschild geometry
defined by a source of mass $\mu$ the gravitational field is weak if
the parameter $G_N\mu L^{-1}$ is small.  In de Sitter geometry with
cosmological constant $\Lambda$ the corresponding parameter is
$\Lambda L^2$.  In the microscopic domain we have two length scales
determined by respectively the square $G_N\hbar$ of the Planck length
and by $\kbar$. These two scales are not necessarily related, although
both are of course much smaller than $L^2$.  It would be reasonable to
identify $\kbar$ with $G_N\hbar$ thus presumably identifying quantum
gravity with noncommutativity; a weaker assumption, that
noncommutativity gives just an effective description of quantum
gravity, would correspond to a vague inequality $ G_N\hbar \simeq \kbar$.

One can also compare a classical gravitational field with
noncommutativity. As noted, the gravitational field is weak if the
dimensionless parameter $\epsilon_{GF} = (G_N \mu)^2 L^{-2}$ is small;
on the other hand, the space-time is almost commutative if the
dimensionless parameter $\epsilon_{NC}= \kbar L^{-2}$ is small.  If
noncommutativity is not related to gravity then it makes sense to
speak of ordinary gravity as the limit $\kbar \to 0$ with $G_N\mu$ non
vanishing. However one could assume that noncommutativity and gravity
are directly related: in that case, both should vanish with $\kbar$.
We shall consider the first situation with
\begin{equation}
 \kbar \ll (G_N\mu)^2 \ll L^2, \qquad
  \epsilon=\epsilon_{NC} /\epsilon_{GF} \ll 1 ,  \label{ekm}
\end{equation} 
and expand in the parameter $\epsilon$ which is a measure of the
relative dimension of a typical `space-time cell' compared with a
typical `quantity of gravitational energy'.

\subsection{Differential calculi}

Assume that there is over $\c{A}$ a differential calculus which is
such~\cite{Mad00c} that the module of 1-forms is free as a left or
right module and possesses a preferred basis $\theta^\alpha$ which
commutes
\begin{equation}
[x^\mu, \theta^\alpha] = 0                            \label{mod}
\end{equation}
with the algebra.  Such a basis we call a frame. The space which one
obtains in the commutative limit is therefore parallelizable, with a
global moving frame $\t{\theta}^\alpha$ which is the commutative
limit of $\theta^\alpha$.  We can write the differential
\begin{equation}
dx^\mu = e_\alpha^\mu \theta^\alpha, \qquad e_\alpha^\mu = e_\alpha x^\mu.
                                                                          \label{teta}
\end{equation}
The algebra is defined  by  product (\ref{xx}) that is by the matrix
of elements $J^{\mu\nu}$; the metric is defined we shall see below
as in the commutative case, 
by the matrix of elements $e_\alpha^\mu$.  Consistency requirements
impose relations between
these two matrices which in simple situations allow us to find a
one-to-one correspondence between the structure of the algebra and the
metric.

The input of which we shall make the most use is the Leibniz rule
\begin{equation}
i\kbar e_\alpha J^{\mu\nu} =
[e^\mu_\alpha, x^\nu] - [e^\nu_\alpha, x^\mu].      \label{lr}
\end{equation}
One can see here a differential equation for $J^{\mu\nu}$ in terms of
$e^\mu_\alpha$.   The momenta $p_\alpha$ introduced  as in quantum mechanics
stand in duality to the position operators $x^\mu$ by the relation
\begin{equation}
[p_\alpha, x^\mu] = \hbar e_\alpha x^\mu = \hbar  e^\mu_\alpha.             \label{pxe}
\end{equation}
The right-hand side of this identity defines the gravitational
field. The left-hand side must obey Jacobi identities. These
identities yield relations between quantum mechanics in the given
space-time and the noncommutative structure of the
algebra. 

The three aspects of reality then, the curvature of space-time,
quantum mechanics and the noncommutative structure  are
intimately connected. We shall explore here an even more exotic
possibility that the field equations of general relativity are encoded
in the structure of the algebra so that the relation between
general relativity and quantum mechanics can be understood by the
relation which each of these theories has with noncommutative
geometry.

In spite of the rather lengthy formalism the basic idea is simple. We
start with a classical geometry described by a moving frame
$\t{\theta}^\alpha$ and we quantize it by replacing the moving frame
by a frame $\theta^\alpha$, as we shall describe in some detail
below. The easiest cases would include those frames which could be
quantized without ordering problems. Let $\t{e}_\alpha$ be the vector
fields dual to the frame $\t{\theta}^\alpha$; we quantize them as
in~(\ref{pxe}) by imposing the rule
\begin{equation}
\t{e}_\alpha \mapsto e_\alpha = \hbar^{-1} \ad p_\alpha .
\end{equation}
Finally, we must construct a noncommutative algebra consistent with
the assumed differential calculus; this defines the image of the map
\begin{equation}
 {\theta}^\alpha \longrightarrow  \t{\theta}^\alpha \longrightarrow J^{\mu\nu}.                   \label{map}
\end{equation}
More details of this map will be given in the Appendix. The algebra we
identify with `position space'. 

 To construct phase space we must add
the momenta $p_\alpha$. In ordinary geometry there is but one way to
do so; the derivations $e_\alpha$ are outer and the associated momenta
do not belong to the algebra generated by the position variables.
That is, we add the extra elements which are necessary in order that
the derivations be inner, as one does in ordinary quantum mechanics. 
In noncommutative geometry there are more possibilities; in particular, 
$p_\alpha$ can belong to the initial algebra $\c{A}$. Also the
 new element  is that the consistency relations in the algebra 
such as Jacobi identities
\begin{equation}
i\kbar [p_\alpha, J^{\mu\nu}] = [x^{[\mu},[p_\alpha,x^{\nu]}]].    \label{Jg}
\end{equation}
in principle restrict $\theta^\alpha$ and $J^{\mu\nu}$.

If the space is flat,  $e^\mu_\alpha = \delta^\mu_\alpha$
and the frame is the canonical flat frame then
the right-hand side of~(\ref{Jg}) vanishes and it is possible to
consistently choose the expression $J^{\mu\nu}$ to be equal to a
constant.  On the other hand, if the space is curved the
right-hand side cannot vanish except of course in the
limit $\kbar\to 0$. The map~(\ref{map}) is not single
valued since any constant $J$ has flat space as inverse image.

We must  insure also  that the differential is well defined. A
necessary condition is that $d[x^\mu, \theta^\alpha] = 0$, from which
it follows that the momenta $p_\alpha$ must satisfy the consistency
condition
\begin{equation}
2 p_\gamma p_\delta P^{\gamma\delta}{}_{\alpha\beta} -
p_\gamma F^\gamma{}_{\alpha\beta} -
K_{\alpha\beta} = 0.                                  \label{consis}
\end{equation}
The $P^{\gamma\delta}{}_{\alpha\beta}$ define the exterior product in the
algebra of forms,
\begin{equation}
 \theta^\gamma \theta^\delta =P^{\gamma\delta}{}_{\alpha\beta}\theta^\alpha
\otimes\theta^\beta.                                      \label{wedge}
\end{equation}
We write $P^{\alpha\beta}{}_{\gamma\delta}$ in the form
\begin{equation}
P^{\alpha\beta}{}^{\phantom{\alpha\beta}}_{\gamma\delta}
= \tfrac 12 \delta^{[\alpha}_\gamma \delta^{\beta]}_\delta
+ \ep Q^{\alpha\beta}{}^{\phantom{\alpha\beta}}_{\gamma\delta}   \label{P}
\end{equation}
of a standard antisymmetrizer plus a peturbation.  If
further~\cite{MacMad03} we decompose
$Q^{\alpha\beta}{}_{\gamma\delta}$ as the sum of two terms
\begin{equation}
Q^{\alpha\beta}{}_{\gamma\delta} =
Q_-^{\alpha\beta}{}_{\gamma\delta} +
Q_+^{\alpha\beta}{}_{\gamma\delta}
\end{equation}
symmetric (antisymmetric) and antisymmetric (symmetric) with
respect to the upper (lower) indices then the condition that
$P^{\alpha\beta}{}_{\gamma\delta}$ be a projector is satisfied to
first order in $\kbar$ because of the property that
\begin{equation}
Q^{\alpha\beta}{}_{\gamma\delta} = P^{\alpha\beta}{}_{\zeta\eta}
Q^{\zeta\eta}{}_{\gamma\delta} + Q^{\alpha\beta}{}_{\zeta\eta}
P^{\zeta\eta}{}_{\gamma\delta}.
\end{equation}
The compatibility condition
\begin{equation}
(P^{\alpha\beta}{}_{\zeta\eta})^* P^{\eta\zeta}{}_{\gamma\delta} =
P^{\beta\alpha}{}_{\gamma\delta}
\end{equation}
 with the product is satisfied provided 
$Q^{\alpha\beta}{}_{\gamma\delta}$ is real.

From~(\ref{mod}) it follows that
\begin{equation}
d[x^{\mu}, \theta^{\alpha}] =
[dx^{\mu}, \theta^{\alpha}] + [x^\mu, d\theta^{\alpha}] =
e^\mu_\beta [\theta^{\beta}, \theta^{\alpha}] -
\tfrac 12 [x^\mu, C^\alpha{}_{\beta\gamma}] \theta^{\beta}\theta^{\gamma}=0.
\end{equation}
We have here introduced the Ricci rotation coefficients
$C^\alpha{}_{\beta\gamma}$.
We find then that multiplication of 1-forms must satisfy
\begin{equation}
[\theta^{\alpha}, \theta^{\beta}] =
\tfrac 12 \theta^\beta_\mu [x^\mu, C^\alpha{}_{\gamma\delta}]
\theta^{\gamma}\theta^{\delta}.                               \label{bffp}
\end{equation}
Consistency requires then that
\begin{equation}
\theta^{[\beta}_\mu
[x^\mu, C^{\alpha]}{}_{\gamma\delta}] = 0 ;                   \label{xC}
\end{equation}
because of the condition~(\ref{mod}) consistency also requires that
\begin{equation}
\theta^{(\alpha}_\mu [x^\mu, C^{\beta)}{}_{\gamma\delta}] =
Q_-^{\alpha\beta}{}_{\gamma\delta}.                              \label{jam}
\end{equation}

We have in general four consistency relations which must be satisfied
in order to obtain a noncommutative extension
of a commutative manifold. They are the Leibniz
rule~(\ref{lr}), the Jacobi identity and the conditions~(\ref{xC})
and~(\ref{jam}) on the differential.  The first two constraints are not 
completely independent of the differential calculus since one involves 
the momentum operators. The condition~(\ref{jam}) follows in general 
from the expression~\cite{Mad00b}
\begin{equation}
C^{\alpha}{}_{\beta\gamma} = F^{\alpha}{}_{\beta\gamma}
-4\ep p_\delta Q_-^{\alpha\delta}{}_{\beta\gamma}  \label{Cp}
\end{equation}
for the rotation coefficients. It follows also from general
considerations that the rotation coefficients must satisfy the gauge
condition
\begin{equation}
e_\alpha C^{\alpha}{}_{\beta\gamma} =  0       .            \label{gauge-co}
\end{equation}
We shall refer to all these conditions as the Jacobi constraints.

\initiate
\section{Phase space}                           \label{np}

The classical phase space associated to Minkowski space-time has eight
dimensions. The associated algebra $\c{A}$ can be considered as the
algebra with four position generators $\t{x}^\mu$ and four momentum generators
$\t{p}_{\alpha}$ subject to the sole relation that they commute. The
quantized phase space is the same but with the Heisenberg commutation
relations
\begin{equation}
[p_{\alpha}, x^\mu] = \hbar \delta^\mu_\alpha           .
\end{equation}
  Over this algebra there is a natural moving frame
$\theta^\alpha = \delta^\alpha_\mu dx^\mu$ the dual derivations of
which are given by
\begin{equation}
e_\beta = \delta_\beta^\nu \p_\nu = \hbar^{-1} \ad p_\beta .
\end{equation}
We shall set $\hbar = 1$ so that the momenta are normalized as we wish
them to be, with $dx^\mu(e_\beta) = \delta^\mu_\beta$.  The moving frame
$\theta^\alpha$ is clearly a frame in the sense we have defined it.
As in the commutative case, since the frame is exact the curvature
vanishes: the associated geometry is flat. 
We have thus a map
\begin{equation}
\mbox{Minkowski} \mapsto  \mbox{Heisenberg}           \label{M2H}
\end{equation}
the extension of which to a geometry with generic curvature we wish to
construct. There exists the commutative limit as special case; the map
takes each 4-geometry into a subalgebra of the algebra of sections of
the form bundle defined by the moving frame. We shall extend this map
to one which includes the corrections of first order in the
noncommutativity parameter $\kbar$. We shall see that the existence of
the extension imposes integrability conditions on the classical limit
of the map. These conditions we conjecture (guess) replace the field
equations. An analogous situation exists in classical gravity when one
attempts to find solutions as perturbation expansions in the
gravitational coupling constant: the existence for example of the
first-order corrections implies that the sources obey the conservation
laws of the flat space-time.

\subsection{Momenta and representations}                \label{mr}

To make transparent the dual roles played by the position and the
momenta we introduce the index $i = (\lambda,\;\alpha)$ and
write a point in phase space as $y^i = (x^\lambda,\;p_\alpha)$. We
lower the index with the metric components 
$g_{ij} = (g_{\mu\nu},\;g^{\alpha\beta})$.  The classical Heisenberg 
commutation relations can be written then as
\begin{equation}
[y^i, y^j] = J^{ij},
\end{equation}
with
\begin{equation}
J^{ij} =
\left(\begin{array}{cc}
0&\delta^{\mu}_{\beta}\\[6pt]
-\delta_{\alpha}^{\nu} &0
\end{array}\right)      .
\end{equation}
The matrix $J$ contains all the information of the system. The even
elements~$J^+$ describe the algebra and the noncommutative
differential calculus, the odd elements~$J^-$ depend directly on the
frame which in turn determines the  limiting
commutative (de~Rham) differential
calculus. The extension of the map~(\ref{M2H}) is equivalent then to a
map
\begin{equation}
J^+ \mapsto J^- .         \label{mapsto}
\end{equation}
The diagonal elements~$J^+$ consist of the six position commutators
\begin{equation} 
[x^\mu, x^\nu] = \kb J^{\mu\nu}                  \label{xx0}
\end{equation} 
as well as of the `dual' momentum commutators 
\begin{equation}
[p_\alpha, p_\beta] = K_{\alpha\beta} +F^\gamma{}_{\alpha\beta}p_\gamma 
- 2\ep Q^{\gamma\delta}{}_{\alpha\beta}p_\gamma p_\delta.      % \label{d20}
\end{equation}
which measure the curvature. There seems to be no obvious property
which would characterize the map (\ref{mapsto}). As solution to a
differential equation it is non-local. Even in the semi-classical
limit it has no obvious characterization.
  
The unusual new feature of the noncommutative extension is the 
possibility that the momenta  which are functions of coordinates
$p_\alpha(x^\sigma)$ and  satisfy all the constraints  exist. In the
commutative limit this would correspond to a section of the frame
bundle; locally at least there are many. We shall refer to the
$p_\alpha(x^\sigma)$ as a `section' of the noncommutative frame bundle
although we have not defined the latter.  If represented by operators
on some Hilbert space, the representation will be irreducible; the
commutant will reduce to the identity. For example if the matrix $J$
is constant and invertable then the elements $\pi_\alpha$ defined by
\begin{equation}
\pi_\alpha = p_\alpha -  p_{0\alpha},  \qquad 
p_{0\alpha} = - (\kb J)^{-1}_{\alpha\mu} x^\mu                  \label{cD}
\end{equation} 
commute with all the position generators. They can then be set to
zero; the momenta and the position operators generate then the same
algebra. We shall assume this case to be generic. We assume that is
there be a solution $p_{0\alpha}(x^\mu)$ to the equations
\begin{equation}
[p_{0\alpha}, x^\mu] = e^\mu_\alpha.
\end{equation}
But if we have independent momenta $p_\alpha$ then also 
\begin{equation}
[p_\alpha, x^\mu] = e^\mu_\alpha.
\end{equation}
We have then two solutions and the differences 
$\pi_\alpha= p_\alpha - p_{0\alpha}$ commute
with the generators $x^\mu$. 

It would be natural to set $\pi_\alpha=0$ to obtain an irreducible
representation of the position algebra; this would lead however to a
manifestly singular commutative limit. We resolve this by requiring
only that the `phase algebra' $\c{T}$ be irreducible.  The projection
of the tangent bundle onto the manifold is in the algebraic
transcription an injection of $\c{A}$ into $\c{T}$. The $p_\alpha$ then are
decomposed as a sum of a section $p_{0\alpha}$ of this bundle and a remainder
$\pi_\alpha$. The subalgebra $\c{A}^\prime$ of $\c{T}$ generated by the $\pi$
commutes with $\c{A}$. The condition we impose is
\begin{equation}
\c{T} = \c{A} \otimes \c{A}^\prime .
\end{equation}
This is certainly true in the commutative limit. The conditions we
have placed on the manifold imply that the tangent bundle is trivial;
the condition could be considered as the statement that this remains
so in the non-commutative extension.

We can write the relation~(\ref{cD}) as the definition of a `covariant
momentum'
\begin{equation}
\pi_\alpha = p_\alpha +  Z_\alpha(x^\mu),\qquad 
Z_\alpha= - p_{0\alpha}                                   \label{pH}
\end{equation}
This is somewhat analogous to gauge transformations; a covariant
derivative is constructed to commute with them.

\subsection{The correspondence}

We shall here consider phase space with eight generators, the position
generators $x^\mu$ and the momenta $p_\alpha$ defining the exterior
derivations. The latter we suppose admits to first order a bracket of the form
\begin{equation}
[p_\alpha, p_\beta] = C^\gamma{}_{\alpha\beta} p_\gamma + K_{\alpha\beta}.
\end{equation}  
This is a central extension of the classical relation satisfied by the derivations.
In general the relation is given by~(\ref{consis}). The center is
non-trivial and is generated by the elements $\pi_\alpha$. 

The rotation coefficients are directly related to the
commutators of the momentum generators. We have seen that the former
are given by the expression~(\ref{Cp}) and the latter
by~(\ref{consis}), which we write in the form
\begin{equation}
[p_\alpha, p_\beta] = \dfrac{1}{\kb} L_{\alpha\beta} 
\end{equation}
with
\begin{equation}
L_{\alpha\beta} = K_{\alpha\beta} + \kb F^\gamma{}_{\alpha\beta} p_\gamma
- 2(\kb)^2 \mu^2 Q^{\gamma\delta}{}_{\alpha\beta} p_\gamma p_\delta  . \label{consis2}
\end{equation}
\begin{equation}
[p_\alpha, x^\mu] = e^\mu_\alpha
\end{equation}
There is therefore a direct connection between the rotation
coefficients and the commutators $J^{\mu\nu}$.  This relation can be
derived directly without explicitly referring to the momenta.

It is easy to see that the Jacobi identity
\begin{align}
&[x^\nu,[p_\alpha,p_\beta]]
+[p_\alpha,[p_\beta,x^\nu]]
+[p_\beta,[x^\nu,p_\alpha]] = 0
\end{align}
is in fact an identity.  Consider the Jacobi identity
\begin{align}
&[p_\alpha,[x^\mu,x^\nu]]
+[x^\mu,[x^\nu,p_\alpha]]
+[x^\nu,[p_\alpha,x^\mu]] = 0 .
\end{align}
It can be written as
\begin{align}
\kb [p_\alpha,J^{\mu\nu}]
-[x^{[\mu}, e^{\nu]}_\alpha] = 0,
\end{align}
from which one derives in the semi-classical approximation a
differential equation
\begin{align}
&e_\alpha J^{\mu\nu}
- J^{[\mu\rho} \p_\rho e^{\nu]}_\alpha = 0.
\end{align}
This condition can be expressed uniquely in terms of frame components
using the sequence of identities
\begin{align}
\theta^\beta_\mu \theta^\gamma_\nu (e_\alpha J^{\mu\nu}- J^{[\mu\rho} \p_\rho e^{\nu]}_\alpha)
&=e_\alpha J^{\beta\gamma} - J^{\mu\nu} e_\alpha(\theta^\beta_\mu \theta^\gamma_\nu)
-J^{[\beta\delta} [p_\delta,[p_\alpha, x^\nu]]\theta^\gamma_\nu
\nonumber\\[4pt]
&=e_\alpha J^{\beta\gamma} - J^{\mu\nu} e_\alpha(\theta^\beta_\mu \theta^\gamma_\nu)  
- J^{[\beta\delta} e_\alpha e^{\nu}_\delta \theta^{\gamma]}_\nu
+ J^{[\beta\delta} [x^{\nu},[p_\delta,p_\alpha]]\theta^{\gamma]}_\nu
\end{align}
which follow immediately from the Leibniz rule.
We find then the condition
\begin{equation}
e_\alpha J^{\beta\gamma} 
+ C^{[\beta}{}_{\alpha\delta} J^{\delta\gamma]} =0.                     \label{1a}
\end{equation}                         
The extension of the map~(\ref{M2H}) which one could propose is
therefore a map
\begin{equation}
e^\mu_\alpha   \mapsto  J^{\mu\nu}
\end{equation}
which one can consider as a map
\begin{equation} 
C^\gamma{}_{\alpha\beta}  \mapsto J^{\alpha\beta}                 \label{MM}
\end{equation}
obtained by solving~(\ref{1a}).

\subsection{Metrics and connections}

We have defined a notion of antisymmetry by the array
$P^{\alpha\delta}{}_{\gamma\beta}$. To define symmetry we introduce a
flip $\sigma$ which exchanges in a twisted way the two factors of a
tensor product. In terms of the frame it can be written
\begin{equation}
\sigma(\theta^\alpha \otimes \theta^\delta) 
= S^{\alpha\delta}{}_{\gamma\beta} \theta^\gamma \otimes \theta^\beta.
\end{equation}
If we require that the map be bilinear then the coefficients must be
constant.  The relation between $P^{\alpha\delta}{}_{\gamma\beta}$ and
$S^{\alpha\delta}{}_{\gamma\beta}$ is the condition
\begin{equation}
\pi \circ (1 + \sigma) = 0,
\end{equation}
the antisymmetric part of a symmetrized tensor should vanish. The
conditions satisfied by the flip are quite simple in the first
approximation we are here considering. If we write
\begin{equation}
S^{\alpha\beta}{}_{\gamma\delta} 
= \delta_\gamma^\beta \delta_\delta^\alpha + \ep T^{\alpha\beta}{}_{\gamma\delta},
\end{equation}
then we find that
\begin{equation}
Q_+^{\alpha\beta}{}_{\gamma\delta} + T^{[\alpha\beta]}{}_{\gamma\delta} = 0.
\end{equation}
Some further relations are given in the Appendix.

With the frame one can construct a metric just as one does in the
commutative case. It is the bilinear map of the tensor product of the
module of 1-forms  by itself into the algebra. It associates therefore
a function to each pair of vector fields. We consider this metric in
the classical approximation. It is defined by the frame and a set of
coefficients, necessarily in the center, by the expression
\begin{equation}
g^{\alpha\beta} = g(\theta^\alpha \otimes \theta^\beta).
\end{equation}
We choose the frame to be orthonormal in the commutative limit; we can
write therefore
\begin{equation}
g^{\alpha\beta} = \eta^{\alpha\beta} - \ep h^{\alpha\beta}.
\end{equation}
In the linear approximation, the condition of the reality of the
metric becomes
\begin{equation}
h^{\alpha\beta} + \bar{h}^{\alpha\beta} 
= T^{\beta\alpha}{}_{\gamma\delta}\eta^{\gamma\delta}.
\end{equation}

The covariant derivative is given by
\begin{equation}
D \xi = \sigma(\xi \otimes \theta) - \theta \otimes \xi .
\end{equation}
In particular
\begin{equation}
D \theta^\alpha = -\omega^\alpha{}_\gamma\otimes\theta^\gamma 
= - (S^{\alpha\beta}{}_{\gamma\delta} -
\delta^\beta_\gamma \delta^\alpha_\delta) 
p_\beta \theta^\gamma \otimes \theta^\delta =
-\ep T^{\alpha\beta}{}_{\gamma\delta}  
p_\beta \theta^\gamma \otimes \theta^\delta ,
\end{equation}
so the connection-form coefficients are linear in the momenta
\begin{equation}
\omega^\alpha{}_\gamma = \omega^\alpha{}_{\beta\gamma}\theta^\beta =
\ep p_\delta T^{\alpha\delta}{}_{\beta\gamma}\theta^\beta .    \label{LeviC}
\end{equation}
On the left-hand side of the last equation is a quantity
$\omega^\alpha{}_\gamma$ which measures the variation of the metric;
on the right-hand side is the array $T^{\alpha\delta}{}_{\beta\gamma}$
which is directly related to the anti-commutation rules for the
1-forms, and more importantly the momenta $p_\delta$ which define the
frame. As $\kbar \to 0$ the right-hand side remains finite and
\begin{equation}
\omega^\alpha{}_{\gamma} \to \t{\omega}^\alpha{}_{\gamma}.
\end{equation}

The connection is torsion-free if the components satisfy the constraint
\begin{equation}
\omega^\alpha{}_{\eta\delta}P^{\eta\delta}{}_{\beta\gamma} 
= \tfrac 12 C^\alpha{}_{\beta\gamma}.          \label{tor-free}
\end{equation}
The connection is metric if
\begin{equation}
\omega^{\alpha}{}_{\beta\gamma} g^{\gamma\delta} +
\omega^{\delta}{}_{\gamma\eta} S^{\alpha\gamma}{}_{\beta\zeta}
 g^{\zeta\eta} = 0,                                      %\label{wkbm-c3}
\end{equation}
or linearized,
\begin{equation}
T^{(\alpha\gamma}{}_\delta{}^{\beta)} = 0.                      %\label{comp}
\end{equation}
The equation can be solved, so in the linear approximation every metric has
 associated to it a unique torsion-free metric connection.

\initiate
\section{The perturbation expansion}

We must now examine the conditions under which the noncommutative
frame can be considered as having the classical frame as limit.  By
classical we refer here to ordinary quantum mechanics.  We have three
commutators, position space, momentum space and the cross terms, given
respectively by
\begin{align}
[x^\mu, x^\nu] &= \kb J^{\mu\nu}
\\[4pt]
[p_\alpha, p_\beta] &= \dfrac{1}{\kb} L_{\alpha\beta} 
\\[4pt]
[p_\alpha, x^\mu] &= e^\mu_\alpha
\end{align}
with $L_{\alpha\beta}$ given by Equation~(\ref{consis2}).

\subsection{General quasi-classical limit}                    % \label{mod}

Let us discuss in more detail the limit of the 
frame geometry. Recall the definitions
\begin{equation}
f \theta^\alpha = \theta^\alpha f,\qquad
\theta^\alpha = \theta^\alpha_\mu (x^\sigma) dx^\mu ,
\end{equation}
the second being the inverse of (\ref{teta}). As
functions of coordinates are given through 
the Taylor expansion, the commutator $ [x^\lambda, f(x^\sigma)]$
can be expressed in terms of the basic commutators $J^{\lambda\mu}$. 
Neglecting the operator ordering that is in linear order in $\kbar$ we obtain 
\begin{equation}
[x^\lambda, f(x^\sigma)] = i\kbar J^{\lambda\mu} \partial_\mu f
= i\kbar J^{\lambda\alpha} e_\alpha f .                      \label{::}
\end{equation}
This is the quasi-classical approximation. We have denoted
\begin{equation}
J^{\lambda\mu}= J^{\alpha\beta} e^\lambda_\alpha e^\mu_\beta .
\end{equation}
In particular, 
\begin{equation}
[x^\lambda,dx^\mu]= 
- e^\mu_\alpha [x^\lambda, \theta^\alpha_\nu] dx^\nu =
- i\kbar J^{\lambda\beta} e^\mu_\alpha e_\beta \theta^\alpha_\nu dx^\nu =
 i\kbar J^{\lambda\beta} e_\beta e^\mu_\alpha \theta^\alpha .
\end{equation}
Using the quasi-classical approximation we can obtain an equation for  $J$. From 
\begin{equation}
d J^{\lambda\mu}= dJ^{\alpha\beta} e_\alpha^\lambda e^\mu_\beta +
J^{[\lambda\beta} e_\alpha e^{\mu]}_\beta \theta^\alpha
\end{equation}
and (\ref{::}) we have
\begin{equation}
dJ^{\alpha\beta} + J^{\gamma[\alpha} 
C^{\beta]}{}_{\gamma\delta} \theta^\delta = 0.
\end{equation}
This equation has the integrability condition 
\begin{equation}
d(J^{\gamma[\alpha} 
C^{\beta]}{}_{\gamma\delta} \theta^\delta) = 0,
\end{equation}
that is
\begin{equation}
d(J^{\gamma[\alpha} C^{\beta]}{}_{\gamma\delta}) \theta^\delta +
J^{\gamma[\alpha} C^{\beta]}{}_{\gamma\delta} d\theta^\delta = 0.
\end{equation}
Using the known expression for $dJ^{\alpha\beta}$ and
the Bianchi identities 
\begin{equation}
\epsilon^{\alpha\beta\gamma\delta}(e_\delta C^{\zeta}{}_{\beta\gamma} +
C^{\zeta}{}_{\delta\eta} C^{\eta}{}_{\beta\gamma}) = 0
\end{equation}
one shows that the first term is identically zero. The
condition reduces to
\begin{equation}
J^{\gamma[\alpha} C^{\beta]}{}_{\gamma\delta} d\theta^\delta = 0.  \label{IC}
\end{equation}

\initiate
\section{Conclusion}

We have argued in favor of considering noncommutative geometry as a
deformation of phase space rather than position space. From this point
of view the algebra and the calculus are on the same footing and in
fact one could avoid to a certain extent at least using the later
since the 1-forms have been encoded in the momenta. Derivations can be
almost identified with momenta but only if one neglects an additive
constant in the latter. We have shown that in noncommutative geometry
there is a preferred origin to the momenta, somewhat analogous to the
preferred origin in the module of 1-forms. This is quite consistent
with previous results to the effect that the commutators determine not
only the structure of the algebra but also the metric of the associated 
geometry.

\initiate
\section{Appendices}

\subsection{Gauge dependence}

We have found a map between the symplectic and the metric structures
on a manifold. The definition is valid only in the semi-classical
approximation and relies essentially on the existence of a frame. An
interesting problem is the study of the variation of the map. For
example one might inquire into the type of variations of the frame 
which leave the symplectic structure invariant and inversely. Both of
these variations could be considered as `gauge' transformations. We
have succeeded in solving only partially this problem.

Consider two choices of commutator $J^{\alpha\beta}$ and
$J^{\prime\alpha\beta}$ with
\begin{equation}
J^{\prime\alpha\beta} = J^{\alpha\beta} + \delta J^{\alpha\beta}
\end{equation}
Then the corresponding variation of the rotation coefficients
$\delta C^\gamma{}_{1\alpha\beta}$  is
given by a solution to the constraints
\begin{align}
&
\delta e_\gamma J^{\alpha\beta} + e_\gamma \delta J^{\alpha\beta} 
-\delta C^{[\alpha}{}_{\gamma\delta} J^{\beta]\delta}
- C^{[\alpha}{}_{\gamma\delta} \delta J^{\beta]\delta} = 0,                \label{cl3}
\\[6pt]&
\epsilon_{\alpha\beta\gamma\delta} \delta J^{\alpha\epsilon}
e_\epsilon J^{\beta\gamma} 
+ \epsilon_{\alpha\beta\gamma\delta}J^{\alpha\epsilon}
\delta e_\epsilon J^{\beta\gamma} 
+ \epsilon_{\alpha\beta\gamma\delta} J^{\alpha\epsilon}
e_\epsilon \delta J^{\beta\gamma} = 0,                             \label{mb*3}
\\[6pt]&
[\delta e_\alpha, e_\beta] + [e_\alpha, \delta e_\beta] 
= \delta C^\gamma{}_{\alpha\beta} e_\gamma + C^\gamma{}_{\alpha\beta} \delta e_\gamma.
\end{align}
This can be simplified if we use the momenta. We conclude from~(\ref{consis7})
that a variation of the momenta must satisfy the constraint
\begin{equation}
[\delta p_\alpha, p_\beta] 
+ [p_\alpha, \delta p_\beta] 
= \delta C^\gamma{}_{\alpha\beta} p_\gamma +                     
 C^\gamma{}_{\alpha\beta} \delta p_\gamma .                     
\end{equation}
with
\begin{equation}
\delta C^\gamma{}_{\alpha\beta} = \delta F^\gamma{}_{\alpha\beta} 
- 4\ep \delta Q^{\gamma\delta}{}_{\alpha\beta}p_\delta    .         %    \label{s-e2}
\end{equation}
In particular the rotation coefficients vary only with a change in the
coefficients.

\subsection{Rotation coefficients}

One can show quite generally that the momenta $p_\alpha$ have a bracket of
the form
\begin{equation}
[p_\alpha, p_\beta] = K_{\alpha\beta} +F^\gamma{}_{\alpha\beta}p_\gamma 
- 2\ep Q^{\gamma\delta}{}_{\alpha\beta}p_\gamma p_\delta.       \label{consis7}
\end{equation}
For the formalism to work we must be able to impose the gauge condition
\begin{equation}
E_{\alpha\beta} = 0,\qquad E_{\alpha\beta} = e_\gamma C^\gamma{}_{\alpha\beta}.
\end{equation}
There is yet another stronger condition to be considered below.

Consider the equations
\begin{equation}
[e_\alpha, e_\beta] = C^\gamma{}_{\alpha\beta} e_\gamma  \label{d19}
\end{equation}
To obtain them as the commutative limits of a set of noncommutative
extensions we introduce the commutators
\begin{equation}
[p_\alpha, p_\beta] = X_{\alpha\beta}          .         \label{d20}
\end{equation}
We can consider the left-hand side of~(\ref{d19})as a limit of the
left-hand side of~(\ref{d20}) in the sense that
\begin{equation}
[[p_\alpha, p_\beta],f] = [[e_\alpha, e_\beta], f].
\end{equation}
For the right-hand sides to satisfy such a relation
we require coefficients $K_{\alpha\beta}$, $F^\gamma{}_{\alpha\beta}$,
$Q^{\gamma\delta}{}_{\alpha\beta}$ such that the matrix of quadratic polynomials
\begin{equation}
X_{\alpha\beta} = K_{\alpha\beta} +
F^\gamma{}_{\alpha\beta}p_\gamma 
- 2\ep Q^{\gamma\delta}{}_{\alpha\beta}p_\gamma p_\delta
\end{equation}
have the property that
\begin{equation}
\lim_{\kbar\to 0} [X_{\alpha\beta}, f] =
C^\gamma{}_{\alpha\beta} \lim_{\kbar\to 0}[p_\gamma, f].  \label{jvm}
\end{equation}
and so we must find coefficients $Q^{\gamma\delta}{}_{\alpha\beta}$ as
well as momenta $p_\alpha$ such that
\begin{equation}
C^\gamma{}_{\alpha\beta}\lim_{\kbar\to  0}[p_\gamma,f]
=  F^\gamma{}_{\alpha\beta}\lim_{\kbar\to  0}[p_\gamma,f]
- 4\lim_{\kbar\to 0}\ep Q^{\gamma\delta}{}_{\alpha\beta}p_\delta [p_\gamma, f].
\end{equation}
Since $f$ is arbitrary we must have therefore to lowest order in $\kbar$
\begin{equation}
C^\gamma{}_{\alpha\beta} = F^\gamma{}_{\alpha\beta} 
- 4\ep Q^{\gamma\delta}{}_{\alpha\beta}p_\delta        .         \label{s-e2}
\end{equation}
This is formally and to first order the same as
\begin{equation}
C^\gamma{}_{\alpha\beta} = \dfrac{\p}{\p p_\gamma} X_{\alpha\beta}.
\end{equation}
If we take another derivative we obtain
\begin{equation}
\dfrac{\p}{\p p_\delta} C^\gamma{}_{\alpha\beta} 
= \dfrac{\p^2}{\p p_\delta \p p_\gamma} X_{\alpha\beta}
= -4\ep Q^{\gamma\delta}{}_{\alpha\beta}.
\end{equation}
From this follows the condition
\begin{equation}
\dfrac{\p}{\p p_\delta} C^\gamma{}_{\alpha\beta} 
- \dfrac{\p}{\p p_\gamma} C^\delta{}_{\alpha\beta} = 0.       \label{Darb}
\end{equation}
We see then also that
\begin{equation}
e_{\delta}C^\gamma{}_{\alpha\beta} 
=  X_{\zeta\delta} \dfrac{\p}{\p p_\delta} C^\gamma{}_{\alpha\beta} 
=-4\ep X_{\delta\zeta}  Q^{\gamma\zeta}{}_{\alpha\beta}
\end{equation}
and from this the integrability conditions
\begin{equation}
e_{\gamma}C^\gamma{}_{\alpha\beta} = 0
\end{equation}
If the classical rotation coefficients do not satisfy this condition
for some choice of $p$-variables they cannot be `quantized'.  We see
here that the $p$-variables are somewhat analogous to the special
variables of the classical Darboux theorem. In this case the
transformation to the special coordinates is the Fourier transform
$x^\mu \mapsto p_\alpha$.

\subsection{The cocycle}

We return now to the cocycle condition~(\ref{cocon}).  Either $F = dA$
for some 1-form $A$ or there is no such 1-form. We know of no $F$
which is not of the form $F=dA$ but there is a case with $F=dA_D$
where $A_D$ has no regular commutative limit.  We know that the Dirac
operator $\theta = - p_\alpha \theta^\alpha$ diverges in the
commutative limit and that the limit of
\begin{equation}
A_D = - i\kbar \theta
\end{equation}
is finite but not everywhere well defined.  We notice then that
the square of $\theta$ can be written as
\begin{equation}
\theta^2 = \tfrac12 [p_\alpha, p_\beta]\theta^\alpha \theta^\beta.
\end{equation}
From this it follows that
\begin{align}
&
d\theta + \theta^2 = - \tfrac 12  [p_\alpha, p_\beta] \theta^\alpha \theta^\beta
+ \tfrac 12 p_\gamma C^\gamma{}_{\alpha\beta} \theta^\alpha \theta^\beta
\nonumber\\[4pt]&\phantom{d\theta + \theta^2}
= - \tfrac 12  [p_\alpha, p_\beta] \theta^\alpha \theta^\beta
+ \tfrac 12 p_\gamma F^\gamma{}_{\alpha\beta} \theta^\alpha \theta^\beta
- 2 \ep p_\gamma p_\delta Q^{\gamma\delta}{}_{\alpha\beta}
\nonumber\\[4pt]&\phantom{d\theta + \theta^2}
= - \tfrac 12 ( [p_\alpha, p_\beta]
- p_\gamma F^\gamma{}_{\alpha\beta}
+ 4 \ep p_\gamma p_\delta Q^{\gamma\delta}{}_{\alpha\beta})
\theta^\alpha \theta^\beta
\nonumber\\[4pt]&\phantom{d\theta + \theta^2}
= - K.
\end{align}
where we have set $ K =  \tfrac 12 K_{\alpha\beta}\theta^\alpha \theta^\beta.$
We can conclude then that
\begin{align}
dA_D + A_D^2 = \kb K
\end{align}
Equations~(\ref{1a}) can be written also in terms of the dual
quantities
\begin{equation}
J^*_{\alpha\beta} 
= \tfrac 12 \epsilon_{\alpha\beta\gamma\delta} J^{\gamma\delta}
\end{equation}
as
\begin{align}
&
e_\alpha J^*_{\beta\gamma} +
C^\delta{}_{\alpha[\beta}J^*_{\gamma]\delta}
+ C^\delta{}_{\alpha\delta}J^*_{\beta\gamma} = 0.
\\[6pt]&
C^\alpha{}_{[\alpha\gamma} J^*_{\beta]\delta} J^{\delta\gamma} = 0.
\end{align}
It will be convenient to introduce the suggestive notation
\begin{equation}
F_{\alpha\beta} = (J^{-1})_{\alpha\beta}. 
\end{equation}
We could also have written
\begin{equation}
F_{\alpha\beta} = |J|^{-1}J^*_{\alpha\beta},\qquad 
|J|^2 = \frac 14 J^*_{\alpha\beta} J^{\alpha\beta}.
\end{equation}
We can now rewrite the equations in terms of the inverse.

From Equation~(\ref{1a}) one can derive the identity
\begin{equation}
e_\alpha F_{\beta\gamma} +
F_{\alpha\delta}C^\delta{}_{\beta\gamma} = 0           \label{cll}
\end{equation}
for the derivative of the inverse if it exists.  This can also be
written as a `cocycle condition'
\begin{equation}
 dF = 0                                    \label{cocon}
\end{equation}
if we introduce the 2-form
\begin{equation}
F = \tfrac 12 F_{\alpha\beta} \theta^\alpha \theta^\beta .
\end{equation}
One can solve (\ref{cll}) for the rotation coefficients. One obtains
\begin{equation}
C^\alpha{}_{\beta\gamma}
 = J^{\alpha\eta} e_\eta F_{\beta\gamma}.            \label{mbb}
\end{equation}
It follows that in the quasi-classical approximation, the linear
curvature is a polynomial in the commutator $J$ and its inverse and
their derivatives.

If we consider $F$ as a Maxwell field
strength then there is a source given by
\begin{equation}
e^\alpha F_{\alpha\beta} 
= F{}^{\alpha\gamma} C_{\alpha\beta\gamma}.
\end{equation}
It follows also from the condition~(\ref{gauge-co}) that the commutator
must necessarily satisfy the constraint
\begin{equation}
e_\alpha \left(J^{\alpha\eta}
e_\eta F_{\beta\gamma}\right) = 0.              \label{c}
\end{equation}
This can also be written as
\begin{equation}
(e_\alpha J^{\alpha\zeta}
+ J^{\alpha\eta} C^\zeta{}_{\alpha\eta})\,
 e_\zeta F_{\beta\gamma}=0.
\end{equation}
If we equate the Expression~(\ref{mbb}) for the rotation coefficients
with that in terms of the components of the frame we find after a few
simple applications of the Leibniz rule that
\begin{equation}
(dF)_{\alpha\beta\gamma}
= e^\mu_{[\beta} e_{\gamma]} F_{\alpha\mu.}
\end{equation}
The cocyle condition~(\ref{cocon}) is equivalent to the condition
\begin{equation}
e^\mu_{[\beta} e_{\gamma]} F_{\alpha\mu.} = 0.
\end{equation}
An interesting particular solution is given by constants:
\begin{equation}
F_{\alpha\mu.} =F_{0\alpha\mu.}.
\end{equation}
It follows then that
\begin{equation}
J^{\mu\nu} = J_0^{\mu\alpha}e^\nu_\alpha, \qquad
J_0^{(\mu\alpha}e^{\nu)}_\alpha = 0.
\end{equation}
One verifies that
\begin{equation}
C^\alpha{}_{\alpha\gamma}
= J^{\alpha\eta} e_\eta F_{\alpha\gamma}
= e_\eta J^{\alpha\eta} F_{\alpha\gamma}
\end{equation}
and so the left-hand side vanishes if and only if
\begin{equation}
e_\beta J^{\alpha\beta} = 0.
\end{equation}

\providecommand{\href}[2]{#2}\begingroup\raggedright\endgroup

\end{document}